\documentclass[prb,twocolumn,notitlepage,showpacs,amsmath,amstex,amssymb,citeautoscript,longbibliography,superscriptaddress,groupedaddress,]{revtex4-1}
\pdfoutput=1

\usepackage{natbib}
\usepackage[english]{babel}
\usepackage{letltxmacro}
\usepackage{latexsym}
\LetLtxMacro{\ORIGselectlanguage}{\selectlanguage}
\makeatletter
\DeclareRobustCommand{\selectlanguage}[1]{%
  \@ifundefined{alias@\string#1}
    {\ORIGselectlanguage{#1}}
    {\begingroup\edef\x{\endgroup
       \noexpand\ORIGselectlanguage{\@nameuse{alias@#1}}}\x}%
}
\newcommand{\definelanguagealias}[2]{%
  \@namedef{alias@#1}{#2}%
}
\makeatother
\definelanguagealias{en}{english}
\definelanguagealias{English}{english}
\usepackage{graphicx}
\usepackage{amsmath}
\usepackage{mathtools}
\usepackage{amsfonts}
\usepackage{amssymb,bbding}
\usepackage{bm}
\usepackage{color}
\usepackage[percent]{overpic}
\usepackage{soul} 
\usepackage{amssymb}
\usepackage{wasysym}
\usepackage{dsfont}
\usepackage{float}
\usepackage{braket}
\usepackage{subfiles}
\usepackage{epstopdf}

\usepackage[normalem]{ulem}

\usepackage{tikz}
\usetikzlibrary{positioning}

\usepackage{blkarray}
\usepackage{multirow}

\newcommand{\be}{\begin{equation}}
\newcommand{\ee}{\end{equation}}
\newcommand{\bea}{\begin{eqnarray}}
\newcommand{\eea}{\end{eqnarray}}

\usepackage{tabularx}
\newcolumntype{L}{>{\raggedright\arraybackslash}X}
\usepackage{makecell}
\begin{document}


\title{The threshold effects in one-dimensional strongly-interacting systems out of equilibrium}

\author{Artem Borin and Eugene Sukhorukov}
\affiliation{
 D\'epartement de Physique Th\'eorique, Universit\'e de Gen\`eve - CH-1211 Gen\`eve 4, Switzerland}

\date{\today}

\begin{abstract}
In this work we investigate the phenomena associated with the new thresholds in the spectrum of excitations arising when different one-dimensional strongly interacting systems are  voltage biased and weakly coupled by tunneling. We develop the perturbation theory with respect to tunneling and derive an asymptotic behavior of physical quantities close to threshold energies. We reproduce earlier results for the electron relaxation at the edge of an integer quantum Hall system and for the non-equilibrium Fermi edge singularity phenomenon. In contrast to the previous works, our analysis does not rely on the free-fermionic character of local tunneling, therefore we are able to extend our theory to wider class of systems, without well-defined electron excitations, such as spinless Luttinger liquids and chiral quantum Hall edge states at fractional filling factors. 
\end{abstract}

\maketitle

\section{Introduction}
Interactions between particles is an important component for the realistic description of many-body systems. While in a large class of systems, such as electrons in metals, interactions can either be neglected or considered perturbatively, in many systems of reduced dimensionality they manifest themselves in new physical effects and even in new states of matter. Among them, there are models of particular importance such as Luttinger liquids \cite{tom,*lut} (LL), quantum dots (QD)  demonstrating the Fermi edge singularity (FES) effect, \cite{FES_Noz} and other systems\cite{kon,*ort,*zerob}, which are exactly solvable for  arbitrary interaction strengths. At equilibrium, these systems have a peculiar behavior. For instance,  the tunneling density of states (TDOS) in LLs has a power-law singularity at Fermi level. In the case of FES a similar dependence is observed for the transition rate between an impurity and a Fermi sea as a function of the energy of the impurity level.\nocite{cotun}

It seems to be natural to propose a generalization of these models by introducing new energy scales and additional thresholds in the spectrum. For instance, this can be done by accounting for  a non-linearity of the electron spectrum in the LL model\cite{glaz}. Alternatively, one can inject non-equilibrium electrons to a LL from a metallic reservoir and study their relaxation to a non-trivial stationary state. \citep{levk,mirLL} Yet another example is the FES effect, where an impurity that hosts a virtual electronic level couples two electronic reservoirs with different chemical potentials by means of a cotunneling process\cite{cotun}. 

It turns out that away from equilibrium these phenomena are deeply related. For instance, the electron relaxation at the edge of an integer quantum Hall (QH) system\citep{levk} at filling factor $\nu = 2$ and the non-equilibrium FES effect in a QD embedded in a QH system\citep{cher} \nocite{aban,mirLL} have been solved using the same approach,  which is based on the evaluation of the full counting statistics (FCS)\citep{lev}  of electron tunneling. A different method has been used to address the FES problem in Refs.\ [\onlinecite{mirLL}] and [\onlinecite{aban}], and to find the TDOS in LLs in Ref.\  [\onlinecite{mirLL}], where the quantities of interest are expressed through a non-equilibrium electron Green's function, represented as a Fredholm determinants over single-particle degrees of freedom. Notably, all these methods rely on the free-fermionic character of the injection of electron excitations.

In this paper we present different approach, which on one hand  can reproduce the results mentioned above, and on the other hand, is also applicable to  systems without well-defined electronic excitations. Namely, we consider a stationary TDOS at the edge of a fractional QH system and in the bulk of a LL away from equilibrium. In both cases we study the relaxation of the non-equilibrium state, created by injecting electrons via a quantum point contact (QPC) from a reservoir with the chemical potential $\mu$. We assume weak tunneling coupling at the QPC (with tunneling probability $T\ll 1$) and study tunneling perturbatively. The  correction to the equilibrium TDOS is then measured by tunneling to a QD at the energy $\epsilon$.

Let us point out that even though the perturbative character of the non-equilibrium tunneling is crucial for our analysis, the obtained results are universal. The quantities of interest (TDOS, transition rates) are typically studied close to the threshold energies~\cite{levk,cher,aban,mirLL}, where they have a universal power-law behaviour $|\epsilon-\epsilon_0|^{\kappa}$ as a function of the energy $\epsilon$ in the vicinity of the energy thresholds $\epsilon_0 = 0$ and  $\epsilon_0 = \pm\mu$.  High-order tunneling processes at the source QPC smear out the singularities at energies of the order of $T\mu$  at zero temperature.~\cite{levk,cher,aban,mirLL} Our main goal, however, is to find universal exponents $\kappa$ in different physical situations, which justifies our perturbative approach. 

The results of our calculations are summarized in Tab.\ \ref{tab:sum} for four different  physical situations. We study the relaxation of a non-equilibrium state at the edge  of a QH system  at the filling factor $\nu=2$. This system has been extensively studied experimentally\citep{pier1,*pier2,*pier3}. 
The measured quantity is the energy dependent correction to the TDOS. We consider the non-equilibrium FES phenomenon in a QH effect based device. This system has been experimentally studied in Ref.\ [\onlinecite{heib}]. The quantity of interest is the sequential tunneling rate as a function of the energy of the QD level. For both of these systems, we reproduce earlier found results, obtained with the non-perturbative methods\citep{levk,cher}. Finally, we evaluate the TDOS in  non-equilibrium LLs and in  fractional QH systems\citep{pier1,*pier2,*pier3,LL1,*LL2,*LL3,*LL4,*LL5,*LL6}.

\begin{table}[h!]
    \label{tab:sum}
  \begin{center}
    \begin{tabularx}{0.9\columnwidth}{|c!{\vrule width 1pt}L|c|L|}
    \hline
     $\epsilon_0$ & $ -\mu+0$ & $ 0$ & $ \mu-0$\\
      \Xhline{1pt}
      $\nu = 2$ & $2(1+\alpha^2)$ & $-1$ & $2(1-\alpha^2)$\\
      \hline
      FES $\Gamma_\pm$ & $2-\alpha_D \pm$\ \ $ 2(1-\eta_D)$ & $-1-\alpha_D$ & $2-\alpha_D \mp$\ \ $ 2(1-\eta_D)$\\
      \hline
      LL & $3\frac{K+K^{-1}}{2} + 1$ & $\frac{K+K^{-1}}{2} - 2$ & $\frac{K+K^{-1}}{2} - 1$\\
      \hline
      $\nu = \frac{1}{2n+1}$ & absent & $0$ & $2n$\\
      \hline
    \end{tabularx}
     \caption{The results for the exponents $\kappa$ of an asymptotic power-law behavior for different quantities considered in the paper are summarized in this table. This includes: (i) TDOS at the edge of a QH system at the filling factor 2 as a function of the dimensionless interaction parameter $\alpha$; (ii) FES sequential tunneling rates to and from a QD embedded in a QH system as a function of the equilibrium FES exponent $\alpha_D$ and of the induced charge $\eta_D$; (iii) TDOS in a spinless LL as a function of the LL parameter $K$; (iv) TDOS at the edge of a chiral fractional QH system at the filling factor $\nu=1/(2n+1)$. In all the cases, this physical quantities are weakly perturbed by injecting non-equilibrium electrons from a metallic systems with the chemical potential $\mu$ and detected at relatively large distances with the help of a QD at the energy $\epsilon$. They show an asymptotic behavior $|\epsilon - \epsilon_0|^{\kappa}$ in the vicinity of different threshold energies $\epsilon_0=-\mu,0,\mu$. }
  \end{center}
\end{table}
\normalsize
The rest of the paper is organized as follows. In Sec.\ \ref{sec:QH} we focus on the physics of electron relaxation at the edge of an integer QH system. We formulate the problem of finding the non-equilibrium correction to the TDOS, develop the tunneling perturbation theory, and find the asymptotic behavior of the correction at different threshold energies. In this section we also recall the essential elements of the  bosonization technique\citep{Gia}. In Sec.\ \ref{sec:fes} we concentrate on the non-equilibrium FES and find exponents of singularities in sequential tunneling rates. Finally, in Sec.\ \ref{sec:non-f} we apply our theory to essentially non-fermionic systems: spinless LLs and chiral fractional QH systems. In the Appendix \ref{sec:det},  we derive the perturbative correction to  TDOS at the integer QH edge
 directly from the Fredholm determinant.

\section{QH system at filling factor $\nu = 2$}
\label{sec:QH}

\subsection{Formulation of the problem}
\begin{figure}[h]
	\centering
		\includegraphics[scale=0.6]{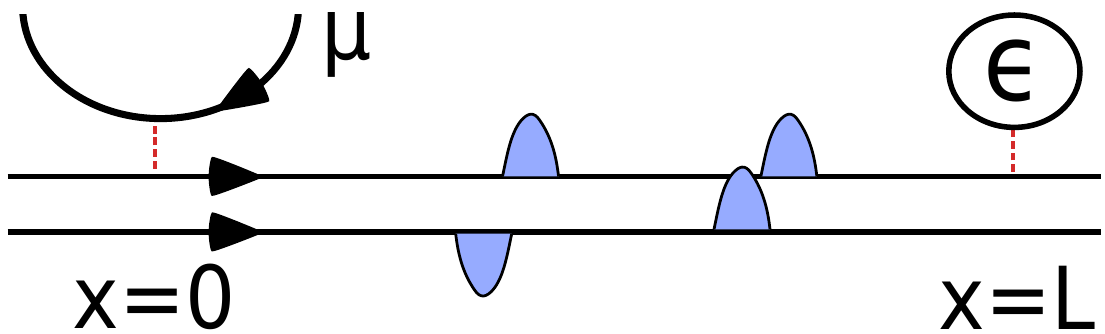} 
	\caption{QH edge states at the  filling factor $\nu=2$ are schematically shown. Due to  strong interactions at the edge, two free-propagating plasmonic modes arise, dipole and charged mode. At the point $x=0$ non-equilibrium electrons are injected from the source channel biased with the chemical potential $\mu$, and at the point $x=L$ the TDOS  $n(\epsilon)$ is measured at the energry $\epsilon$.}
\label{fig:IQH}
\end{figure}

To study a stationary state of the strongly interacting QH edge channels at the filling factor $\nu = 2$ let us consider the system presented in Fig.\ \ref{fig:IQH} that has been realized experimentally in [\onlinecite{pier1,pier2,pier3}]. The dynamics in the interacting channels is governed by the Hamiltonian 
\begin{align}
\label{H0}
H_0&=\pi\int dx\sum_{i}v_{ i}\rho_i^2(x) \nonumber\\
&+\frac{1}{2}\int dxdy\sum_{ij}\rho_i(x)V_{ij}(x,y)\rho_j(y),
\end{align}
where $\rho_i(x)$, $i=U,D$ is the electron density in the upper and lower edge channel, respectively, and  
$V_{ij}(x)$ is the Coulomb potential, which is assumed to be screened at distances smaller than the size of the experimental set-up. Thus, it can be written as $V_{ij}(x) = V_{ij}\delta(x)$. Below, this simplification helps us to diagonalize the Hamiltonian (\ref{H0}) using the bosonization technique, which we recall next.

 We introduce two bosonic fields $\phi_i(x,t)$, $i=U,D$, corresponding to two edge channels, and satisfying the  commutation relations
\begin{equation}
[\partial_x\phi_i(x),\phi_j(y)]=2\pi i\delta_{ij}\delta(x-y).
\end{equation}
Two important identities relate these bosonic fields to edge electrons and charge densities:
\begin{equation}
\label{bos_iden}
\psi_i(x) \propto e^{i\phi_i(x)},\quad \rho_i(x)=\frac{1}{2\pi}\partial_x \phi_i(x),
\end{equation}
where the vertex operator $\psi_i(x) $ annihilates an electron at the point $x$ in the channel $i=U,D$. 
Despite interactions, the edge Hamiltonian (\ref{H0}) is quadratic in bosonic fields (quartic in electrons operators),  and thus can be diagonalized by rotating the basis. A general rotation can be written as 
\begin{equation}\label{rotation}
\begin{cases}
\phi_c = (\alpha \phi_U+\beta\phi_D ), \\
\phi_d = (\beta \phi_U - \alpha \phi_D),
\end{cases}
\end{equation} 
where the constants satisfy the normalization $\alpha^2 + \beta^2 =1$. The new fields describe freely propagating fast charge and slow dipole mode at the edge with velocities   $u_c$ and $u_d$, respectively.  It is important to mention, that typically, because of the long-range character of Coulomb interactions, and due to the fact that $v_i\ll V_{ij}$, the parameters acquire the  universal value $\alpha = \beta = 1/\sqrt{2}$.

We consider the situation, where a non-equilibrium state is created by tunneling processes at the point $x=0$, described by the Hamiltonian
\begin{equation}
\nonumber H_T=\tau \psi_\mu(0)^\dagger \psi_U(0) + {\rm h.c.},
\end{equation}
where $\psi_\mu$ and $\psi_\mu^\dagger$ are the operators for  electrons in the biased source channel (see Fig.\ \ref{fig:IQH}) with $\mu$ denoting the applied bias. For our purposes, it is not necessary to introduce a particular Hamiltonian for electrons in this channel, since the only object we need below is the local correlation function, which we choose to have the free fermionic form
\begin{equation}
\label{biased}
\langle \psi_\mu^\dagger(0,t)\psi_\mu(0,0)\rangle\sim e^{i\mu t}/(it+0).
\end{equation}
This is the case for metallic systems as well as for chiral QH edge channels at integer filling factors. \citep{ff1,ff2}

\begin{figure}[h]
	\centering
		\includegraphics[scale=1.0]{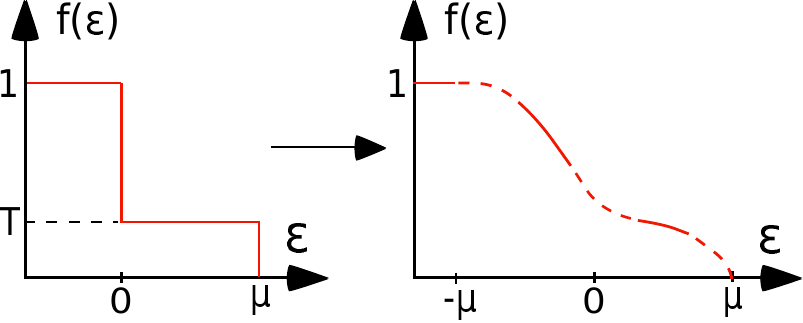} 
	\caption{On the left, the non-equilibrium TDOS is schematically shown directly after the injection from the Fermi sea, biased by the chemical potential $\mu$, to the QH edge at filling factor 2 through a QPC with the transparency $T$. Because of the effectively free-fermionic character of the local tunneling process, the TDOS acquires a well known double-step form. At intermediate distances, due to strong interactions, the double-step TDOS relaxes to a stationary state, as schematically shown on the right. In the regions close to the thresholds (shown by dashed lines) the TDOS acquires a singular power-law behavior, which is the subject of our study.}
\label{fig:thermo}
\end{figure}

At intermediate distances $x = L$ the TDOS $n(\epsilon)$ at the edge reaches a stationary non-equilibrium form (see Fig.\ \ref{fig:thermo}).\cite{levk} 
It can be measured by attaching a QD to the upper edge channel and studying the resonant tunneling current.\cite{pier1,pier2,pier3} We are interested in the deviation of the TDOS from its equilibrium value $n_{eq}(\epsilon)$. It can be presented as following:
\begin{eqnarray}
\label{feps1}
\delta n(\epsilon)&\equiv & n(\epsilon)-n_{eq}(\epsilon) = \int\limits_{-\infty}^\infty dt e^{-i\epsilon t} \delta N(L,t),\\
\label{feps2}
\delta N(L,t)&=&\langle\psi_U^\dagger(L,t)\psi_U(L,0)\rangle_{n-eq}\\ 
&\quad &\quad\quad\quad\quad\quad\quad\quad -\langle\psi_U^\dagger(L,t)\psi_U(L,0)\rangle_{eq}\nonumber ,
\end{eqnarray} 
where the non-equilibrium correlation function is evaluated with respect to the state excited by the source.
The electron operators (\ref{bos_iden}) can be expressed in terms of bosonic eigenmodes by solving equations of motion generated by the Hamiltonian (\ref{H0})
\begin{equation}
\psi_U(x,t) = \exp\left[i\alpha\phi_c(t-x/u_c)+i\beta\phi_d(t-x/u_d)\right],
\end{equation}
where the time dependence reflects the free propagation of eigenmodes with different velocities. 
The first important result that can be easily derived from the bosonic representation is that the local equilibrium TDOS takes the free-fermionic form\citep{ff1,*ff2} in spite of the strong interactions.  At zero temperatur, this gives $n_{eq}(\epsilon)= \theta(-\epsilon)$ and allows one to express $\delta n(\epsilon)$ in terms of the FCS of the free-electron transport.\cite{lev}

In what follows, we rely on  several simplifications for the calculation of the correlation function (\ref{feps2}). First, due to the separation of the spectrum on the charged and dipole mode, propagating with different speeds, one can neglect their correlations at distances $\sim L$, where the stationary state is formed. Indeed, at such distances their contributions to the electron correlation function originate from different tunneling events at $x=0$.  Second, we can ignore the correlations of electrons  that are separated by the distance of the order of $L$.  As a consequence,  only tunneling events that happen at times $\sim- L/u_i$, $i=c,d$, at the point $x=0$ contribute to the correlator in Eq.\ \eqref{feps2}. Finally, we concentrate on the asymptotic forms of the TDOS in order to study its scaling behavior close to Fermi levels. Even though in Secs.\ \ref{sec:fes} and \ref{sec:non-f} we consider different systems, the analysis there can  be also reduced to finding the bosonic correlators in a non-equilibrium state. In the next subsection we show how such quantities can be evaluated.

\subsection{Perturbation theory}
\label{sec:pt}

In order to expand  the correlator  (\ref{feps2}) in powers of the tunneling Hamiltonian  $H_T$, we rewrite it in the interaction representation
\small
\begin{align}
\label{corr}
\delta N(L,t)&=\langle U^\dagger(-\infty,t)\psi_U^\dagger(L,t)U(t,0)\psi_U(L,0)U(-\infty,0) \rangle_{eq} \nonumber \\ &-\langle\psi_U^\dagger(L,t)\psi_U(L,0)\rangle_{eq},
\end{align}
\normalsize 
where $U(t_1,t_2)=\hat{T}\exp[-i\int_{t_2}^{t_1} dt' H_T(t')]$ is the time-ordered evolution operator. 
Expanding the evolution operators up to the second order in $H_T$ generates $3! = 6$ terms. However, the number of terms can be halved. Indeed, for large $L$ tunneling at the point $x = 0$ taking place between times $0$ and $t$ cannot affect results of the measurement at the point $x = L$. Therefore, the evolution operator $U(t,0)$ can be dropped in the above expression. On the physics level, this amounts to neglecting exchange effects in tunneling events at $x=0$ and $x=L$,
 i.e., terms like $\langle\psi_U^\dagger(L,t)\psi_U(0,0)\rangle_{eq}$ are neglected. 

 Moreover, we can safely extend the time domains of the remaining evolution operators to infinity without affecting the correlator \eqref{corr}. Indeed, although by doing so we add extra tunneling events at point $x=0$, they do not  affect the measurements at point $x=L$, since wave packets do not reach this point. Consequently, the equation \eqref{corr} can be rewritten as
\begin{align}
\delta N(L,t)\approx & \langle U^\dagger(-\infty,\infty)\psi_U^\dagger(L,t)\psi_U(L,0)U(-\infty,\infty) \rangle_{eq}\nonumber   \\ \label{corr_modified}
-&\langle\psi_U^\dagger(L,t)\psi_U(L,0)\rangle_{eq}.
\end{align} 
We note, that this approximation is only valid for relatively large energies $\mu$ and $\epsilon$. The corrections to Eq.\ (\ref{corr_modified}) scale as powers of $u_c/[L\ $min$(\mu,\epsilon)]$ with the exponents of the order of 1 (and  exactly 1 for free fermions).

After expanding the evolution operator up to the second order in $H_T$ and expressing the tunneling Hamiltonian in terms of plasmonic eigenmodes,
\begin{equation}
\label{tanham}
H_T = \tau \psi_\mu^\dagger \exp\left[i(\alpha\phi_c+\beta\phi_d)\right] + {\rm h.c.},
\end{equation}
we evaluate the average in Eq.\ \eqref{corr_modified} with respect to the equilibrium state. The result can be expressed in terms of the four-point correlation functions of the following form 
\begin{align}
\label{corK}
&\langle e^{-i\xi\phi_i(t_1)}e^{i\xi\phi_i(t_2)}e^{-i\lambda\phi_i(t-L/u_i)}e^{i\lambda\phi_i(-L/u_i)}\rangle_{eq} \nonumber\\
&=K_{\xi^2}(t_1,t_2)K_{\lambda^2}(t,0)
\frac{K_{\xi\lambda}(t_1,-L/u_i)K_{\xi\lambda}(t_2,t-L/u_i)}{K_{\xi\lambda}(t_1,t-L/u_i)K_{\xi\lambda}(t_2,-L/u_i)},
\end{align}
where the two point correlator
\begin{equation}
\label{K_def}
K_{\gamma}(t_1,t_2)=\langle e^{-i\sqrt{\gamma}\phi_i(t_1)}e^{i\sqrt{\gamma}\phi_i(t_2)}\rangle \propto 
(i(t_1-t_2)+0)^{-\gamma},
\end{equation}
takes the same form for the two eigenmodes $i=c,d$. 

The correlation function (\ref{corK}) has an important property\citep{Artur} that it acquires a non-trivial form only at $t_1$ and $t_2$ close to the flight time of one of the eigenmodes, $\sim -L/u_i$. Therefore, one can split the function \eqref{corr_modified} into two contributions $\delta N(L,t)=\delta N_c(L,t)+\delta N_d(L,t)$ from the charged and dipole mode. The contribution of the charged mode reads
\begin{align}
\label{Kfast}
&\delta N_c(L,t)\propto K_1(t,0)\iint dt_1dt_2e^{i\mu(t_1-t_2)}  \nonumber \\
 &\times \left( K_2(t_1,\!t_2)\frac{K_{\alpha^2}(t_1,\!t)}{K_{\alpha^2}(t_1,\!0)} - {\rm c.c.}\right) \left( \frac{K_{\alpha^2}(t_2,\!0)}{K_{\alpha^2}(t_2,\!t)} - {\rm c.c.} \right).
\end{align}  
The contribution of the dipole mode can be obtained  by  replacing $\alpha\to\beta$.

We have arrived at the expression (\ref{Kfast}) by applying the perturbation expansion directly to the correlation function \eqref{feps2}. Alternatively, one can apply an expansion in tunneling amplitude to the non-perturbative expression for the TDOS in the  form of a Fredholm determinant. This method, presented in the Appendix\ \ref{sec:det}, is based on the free-fermionic character of the local tunneling transport. The advantage of the approach presented in this section is that it can also be used for tunneling to non-Fermi liquid states, as discussed in Sec.\ 
\ref{sec:non-f}.

\subsection{Asymptotic behaviour of TDOS}
\label{thissec}

In this section we evaluate the TDOS \eqref{feps1} asymptotically close to the threshold energies $\epsilon_0= 0$ and $|\epsilon_0|=\mu$ (see Fig.\ \ref{fig:thermo}). Starting with $\mu,\epsilon>0$, the  contribution of the charged mode  $\delta n_c(\epsilon)=\int dt e^{-i\epsilon t} \delta N_c(L,t)$ can be written as 
\begin{align}
\delta n_c(\epsilon)\!\propto\!\int_{-\infty}^\infty\!\!\! dt_-\!\!\int_0^{\infty}\!\!\! dt_+\!\!\int_0^{\infty}\!\!\! dt\frac{e^{-\epsilon( t+t_+)}e^{i(\mu-\epsilon)t_-}}{(it_-  \! +\! t_+ \! + \! t)(t_- \! - \! i0)^2} \nonumber\\ 
\times\frac{(-t_-+it)^{\alpha^2}(t_--it_+)^{\alpha^2}}{(-it_+)^{\alpha^2}(it)^{\alpha^2}},
\label{distr}
\end{align} 
where we changed the variables in the integral  \eqref{Kfast} to  $t_-=t_2-t_1$, $t_+=(t_1+t_2)/2$. 
For $\epsilon\ll\mu$, this expression takes the form
\begin{align}
\delta n_c(\epsilon)\!\!\propto\!\! \int \!\!dt_-\!\!\int_0^\infty\!\!\!\! dt\!\!\int_0^\infty\!\!\!\! dt_+\frac{e^{i\mu t_-}e^{-\epsilon(t+t_+)}}{(t_--i0)^2(t_++t)}\propto\frac{\mu}{\epsilon},
\end{align} 
where we dropped a small prefactor of tunneling probability $T\ll 1$, since we are only interested in a power-law scaling. This results agrees with the findings of Ref. [\onlinecite{levk}]. The dipole contribution $\delta n_d(\epsilon)$ scales in the same way.

We now concentrate on the behavior close to the second threshold in the TDOS: $\mu-\epsilon\ll \epsilon$. We stress, that this threshold arises in the weak tunneling limit, to leading order in tunneling at the source QPC, because the maximum energy that can be injected with  one electron from the source is equal to $\mu$. Consequently, to leading order in tunneling $\delta n(\varepsilon)=0$ for $\varepsilon>\mu$. High-order tunneling processes smear out the singularity. Close to the threshold the charged mode contribution reads 
\begin{align}
\delta n_c(\epsilon)&\propto\int\!\!dt_-\!\!\int_0^\infty\!\!\!\!\!dt\int_0^\infty\!\!\!\!\!dt_+\frac{e^{i\mu t_-}e^{-\epsilon(t+t_+)}}{i(t_--i0)^{3-2\alpha^2}(it)^{\alpha^2}(-it_+)^{\alpha^2}}\nonumber
\\ &\propto\left(\frac{\mu-\epsilon}{\mu}\right)^{2(1-\alpha^2)},
\end{align}
and similar expression is obtained for the dipole mode by replacing $\alpha\to\beta$.
In the case of strong long-range interactions the charge of the tunneling electron equally splits between charged and dipole mode, $\alpha = \beta =1/\sqrt{2}$, which leads to the linear dependence: $\delta n(\epsilon)\propto (\mu-\epsilon)/\mu$.

\begin{table}[h!]
  \begin{center}
    \begin{tabular}{|c!{\vrule width 1pt}c|c|c|}
    \hline
     $\epsilon_0$ & $ -\mu+0$ & $ 0$ & $ \mu-0$\\
     \Xhline{1pt}
      $\kappa$ & $2(1+\alpha^2)$ & $-1$ & $2(1-\alpha^2)$\\
      \hline
    \end{tabular}
        \caption{For electron tunneling to the edge of an integer QH system at filling factor $\nu = 2$ the correction to the equilibrium TDOS acquires the general asymptotic form $\delta n(\epsilon)\propto|\epsilon - \epsilon_0|^{\kappa}$ . The exponents $\kappa$ of this  asymptotic behaviour in the vicinity of different thresholds $\epsilon_0$  are shown, where $\alpha$ is interaction constant.
\label{tab:IQH}}
  \end{center}
\end{table}

In the next step we analyze the hole part of the TDOS $\epsilon<0$. Since the details of the evaluation of the TDOS \eqref{feps1} are the same, we present the results without the derivation (see Tab.\ \ref{tab:IQH}). Finally, we note that for $\mu<0$ the TDOS is immediately obtained by exchanging electrons and holes, and thus  the following identity holds
\begin{equation}
\label{e-h}
\delta n(-\epsilon)|_{\mu\to -\mu}=-\delta n(\epsilon),
\end{equation} 
which can be derived directly from Eq.\ \eqref{Kfast} and is intuitive from the physics perspective.

\section{Fermi edge singularity}
\label{sec:fes}
\begin{figure}[h]
	\centering
		\includegraphics[scale=0.5]{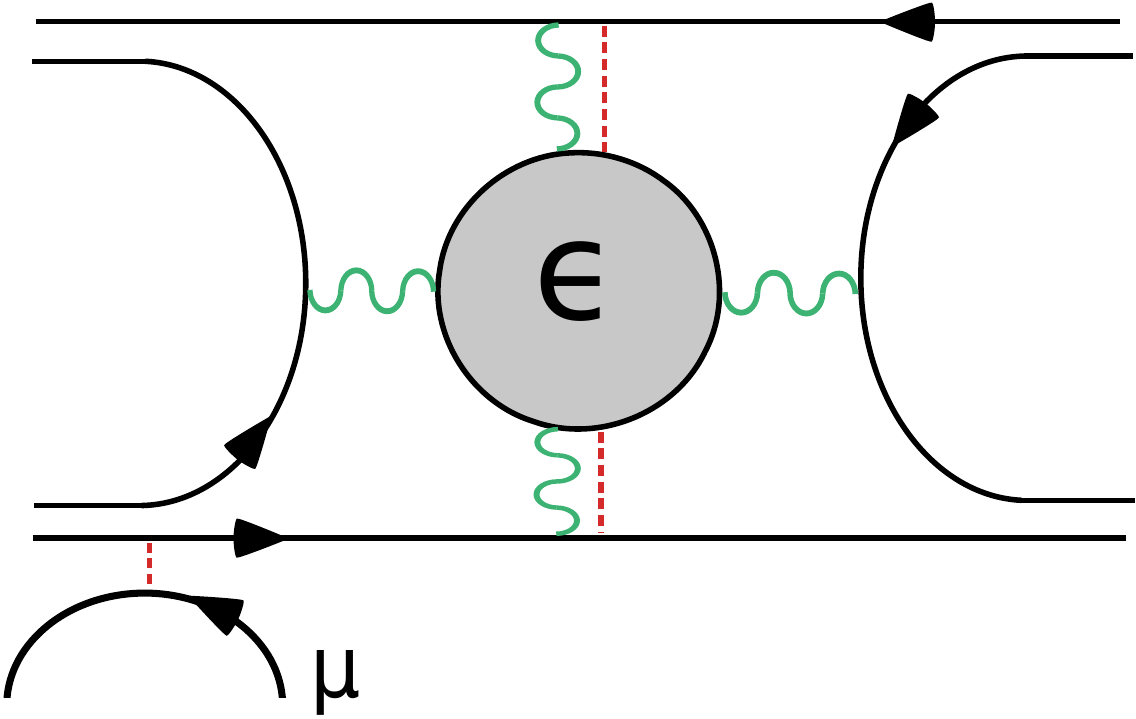} 
	\caption{The QH system at integer filling factor with the embedded QD, perturbed by tunneling at the voltage biased QPC, is schematically shown (for details, see the experiment [\onlinecite{heib}]). The QD strongly interacts with surrounding edge channels (shown by arrows), which partially or completely screen an electron added to the QD energy level $\epsilon$. In equilibrium, this leads to the well-known FES phenomenon: sequential tunneling rates (shown by dashed lines) to and from the energy level $\varepsilon$ (shifted by the interactions, see Eq.\ \ref{renorm}),  acquire the universal low-energy behavior $\Gamma_\pm(\varepsilon)\propto 1/|\varepsilon|^{\alpha_D}$, where the exponent $\alpha_D$ depends only on the charges induced in surrounding channels. Weak tunneling at the upstream QPC, biased with the chemical potential $\mu$, creates new thresholds and modifies FES exponents. They are presented in the table \ref{tab:FES}.}
\label{fig:FES}
\end{figure}

In this section we apply the technique developed above to the problem of non-equilibrium FES. Motivated by the recent experiment,\cite{heib} we study this effect in the QH set-up, shown in the Fig.\ \ref{fig:FES}. In this system,  FES appears as a universal energy dependence of the transition rate between the edge channel and the QD level. We follow the bosonization approach of the paper [\onlinecite{cher}] and present the  Hamiltonian of the system in the form
\begin{equation}
H = H_{0}+ H_{int} +H_T + H'_T,
\end{equation}
where $H_0$ is the free-fermionic part
\begin{equation}
H_0 = \int\frac{dx}{4\pi^2} \sum_i{}v_i(\partial_x\phi_i(x))^2 +\epsilon\, d^\dagger d,
\end{equation}
describing excitations in the QH channels and in the QD, respectively. The summation in the first term runs over four channels surrounding the QD, $i =U,D,L,R$, where $\phi_i(x)$ are bosonic operators introduced in Sec.\ \ref{sec:QH}. The QD is tuned to the resonant tunneling regime via the energy level  $\epsilon$, and operators $d^\dagger$ and $d$ create and annihilate an electron at this level.

The key ingredient of the FES is Coulomb interactions between the charge localized on the QD and the density accumulated in the channels. It is described by the Hamiltonian
\begin{equation}
H_{int} = \frac{1}{2\pi}d^\dagger d \int dx \sum_i U_i(x)\partial_x\phi_i(x),
\end{equation} 
where $U_i(x)$ are the Coulomb potentials, and the sum runs over the surrounding channels, $i=U,D,L,R$. While the general universal solution of the problem can be found
in the paper [\onlinecite{Anya}], we replace potentials with $U_i(x)=U_i\delta(x)$ for simplicity, since we are interested in the low-energy physics, where the length of edge excitations is larger than the range of potentials.\footnote{We note, that the original experiment [\onlinecite{heib}] has been done at the filling factor 2, where each edge contains two channels. However, as explained in the paper [\onlinecite{cher}], interactions in these channels can be neglected, since they do not affect the low-energy physics of FES.}

The upper and lower channel are coupled to the QD through the tunneling Hamiltonian 
\begin{equation}
H_T = d^\dagger\sum_{i=U,D}\tau_i e^{i\phi_i(L)} + {\rm h.c.},
\end{equation}
where $x=L$ is the point at the edges, where tunneling takes place. A non-equilibrium state in the lower channel is created by electron tunneling at point $x=0$ from the source channel, biased with the  chemical potential $\mu$. The Hamiltonian that accounts for this process is given by
\begin{equation}\label{sourceH}
H'_T = \tau\psi_\mu^\dagger(0) e^{i\phi_D(0)} + {\rm h.c.},
\end{equation} 
where operators $\psi_\mu$ and $\psi_\mu^\dagger$ describe electrons in the biased channel. 

According to the paper [\onlinecite{cher}], the bosonization technique allows to treat the interaction term $H_{int}$ exactly. One can perform a unitary transformation that removes this term at the cost of the modification of the energy of the QD level $\epsilon\to\varepsilon$ and of the transformation of the tunneling Hamiltonian $H_T\to \tilde H_T$:
\begin{align}
\varepsilon &= \epsilon + \sum_i \eta_i U_i,\label{renorm} \\ 
\tilde H_T &=  d^\dagger\sum_i \tau_i e^{i\phi_i(L)-\sum_j \eta_j \phi_j(L)}+h.c.,
\end{align}  
where the dimensionless numbers $0\leq \eta_i\leq1$ are the charges accumulated in surrounding channels in response to adding an electron to the QD. For the QD screened solely by these channels, $\sum_i \eta_i = 1$. 

The transition rate from the lower channel to the QD $\Gamma_+$, and the rate for the reversed  process $\Gamma_-$ can be found perturbatively by applying the Fermi golden rule with respect to the modified tunneling Hamiltonian $\tilde H_T$,
\begin{equation}
\label{Gam}
\Gamma_{\pm}(\varepsilon)\propto \int dt e^{- i \varepsilon t}\chi_D(t,\pm(1-\eta_D))\prod_{i\not= D}\chi_i(t,\mp \eta_i),
\end{equation}
where the correlation functions 
\begin{equation}
\label{chi_def}
\chi_i(t,\lambda) = \langle e^{-i\lambda\phi_i(t)}e^{i\lambda\phi_i(0)}\rangle_{n-eq}
\end{equation}
for $i=U, D, L, R$ are evaluated over non-equilibrium state  created by tunneling from the source channel, described by the Hamiltonian (\ref{sourceH}). Since at low energies $H'_T$ perturbs only the lower channel,\footnotemark[\value{footnote}] for other channels, $i\neq D$, the averaging is performed over the equilibrium state, which gives $\chi_i(t,\lambda) =  K_{\lambda^2}(t,0)$ [see Eq.\ \eqref{K_def}]. For the lower channel, averaging has to be evaluated over the non-equilibrium state created by tunneling from the source. We, therefore, apply the perturbation theory, introduced in Sec.\ \ref{sec:QH}.

We skip the details of the calculations, outlined in the Sec.\ \ref{sec:QH}, and present the results for the asymptotic behavior of the transition rates for $\mu>0$  in the table \ref{tab:FES}. The transition rates for the negative bias follow from the electron-hole symmetry, $\Gamma_{\pm}(-\varepsilon)|_{\mu\to-\mu} = \Gamma_{\mp}(\varepsilon)$. These results have to be compared to the well-known FES exponents in equilibrium: $\Gamma_\pm(\varepsilon)\propto 1/ |\varepsilon|^{\alpha_D}$, where $\alpha_D=2\eta_D-\sum_i\eta_i^2$.   Our findings are consistent with those of the papers [\onlinecite{mirLL},\onlinecite{cher}], where the transition rates are evaluated for arbitrary tunneling, so that the singular behaviour at the thresholds acquires the natural cut-off at energies of the order of $\mu$ times the small transparency of the source QPC.
\begin{table}[h!]
    \label{tab:FES}
  \begin{center}
    \begin{tabular}{|c!{\vrule width 1pt}c|c|c|}
    \hline
     $\varepsilon_0$ & $ -\mu+0$ & $ 0$ & $ \mu-0$\\
      \Xhline{1pt}
      $\kappa$  & $2-\alpha_D \pm 2(1-\eta_D)$ & $-1-\alpha_D$ & $2-\alpha_D \mp 2(1-\eta_D)$\\
      \hline
    \end{tabular}
    \caption{The exponents $\kappa$ of the asymptotic behaviour $\Gamma_\pm(\varepsilon)\propto|\varepsilon - \varepsilon_0|^\kappa$ of the transition rates in the vicinity of different thresholds $\varepsilon_0$ are shown. }
  \end{center}
\end{table}

\section{Tunneling to non-Fermi liquids}
\label{sec:non-f}
The systems  considered in the previous sections can be investigated using a non-perturbative method, as discussed in the Appendix\ \ref{sec:det}, which is based on the free-fermionic character of local tunneling process. The goal of this section is to present examples of systems, where the application of the perturbation theory approach developed in Sec.\ \ref{sec:pt} cannot be avoided. Namely, we investigate tunneling transport and a stationary non-equilibrium state in spinless LL and at the edge of a fractional QH system. 

\subsection{Luttinger liquid}
\begin{figure}[h]
	\centering
		\includegraphics[scale=0.6]{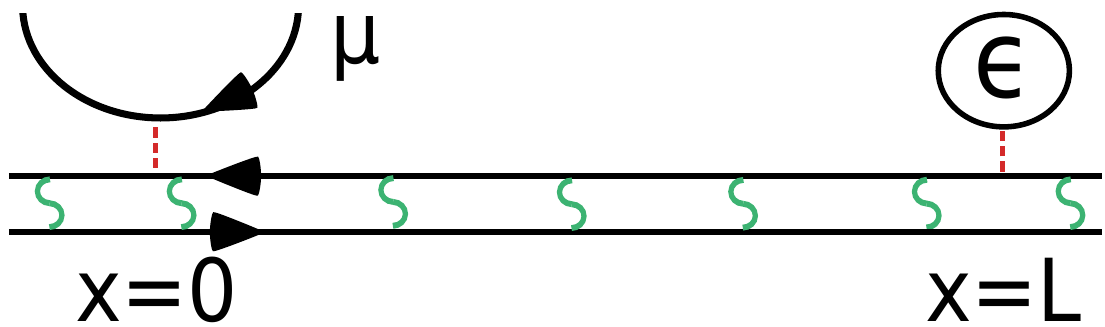} 
	\caption{A spinless LL containing left- and right-moving electrons (presented by arrows) with attached QPC and QD is schematically shown. Wavy lines indicate interactions between electron channels. At the point $x=0$ non-equilibrium electrons are injected from a free-fermionic reservoir with the chemical potential $\mu$. At the point $x=L$ the TDOS $n(\epsilon)$ is measured by a QD.}
\label{fig:LL}
\end{figure}

The interaction induced relaxation in spinless LLs has been studied in Ref.\ [\onlinecite{mirLL}]. However, the analysis in this paper is restricted to the case where a LL is coupled to free-fermionic reservoirs away from the interaction region. This  allows one to reduce the problem to the calculation of a Fredholm determinant of a single-particle operator. Although, the results of this paper can be reproduced with our approach, we go beyond this restriction  and consider a LL system, shown in Fig.\ \ref{fig:LL}, where a non-equilibrium state is created inside LL and interactions cannot be neglected. 
As in the Sec.\ \ref{sec:QH} [see Eqs.\ \eqref{feps1} and \eqref{feps2}], we consider the non-equilibrium correction to the TDOS, that can be measured using resonant tunneling through a QD:
\begin{equation}
\label{fpsi}
\delta n(\epsilon) = \int dt e^{-i\epsilon t} \langle\psi^\dagger(L,t)\psi(L,0)\rangle_{n-eq} - n_{eq}(\epsilon),
\end{equation} 
where $\psi^\dagger$ and $\psi$ are the creation and annihilation electron operators in the LL. Note, that in the case of tunneling to a LL even TDOS $n_{eq}(\epsilon)= \int dt e^{-i\epsilon t} \langle\psi^\dagger(L,t)\psi(L,0)\rangle_{eq}$ has a non-trivial energy dependence.

In a spinless LL, the fermion creation operator $\psi(x,t)$ has two components, $\psi(x,t) = \psi_R(x,t)e^{ik_Fx}+\psi_L(x,t)e^{-ik_Fx}$, that corresponds to the right- and left-moving fermions, where $k_F$ denotes the Fermi wave vector. Right and left movers can be expressed in terms of eigenmodes $\phi_R$ and $\phi_L$ of the LL Hamiltonian, that describe right- and left-moving bosons, respectively\cite{Gia}
\begin{align}
\begin{cases}
\psi_R\propto e^{i(\phi_R \cosh \theta + \phi_L \sinh \theta)},\nonumber\\
\psi_L\propto e^{i(\phi_R \sinh \theta + \phi_L \cosh \theta)},
\end{cases}
\end{align}
where the mixing angle $\theta = \frac{1}{2}\log K$ is determined by the LL interaction parameter $K$. 

A non-equilibrium state in the LL is created by a voltage biased QPC and described  by the tunneling Hamiltonian $H_T$, acting at the point $x=0$:
\begin{equation}
H_T = \tau \psi_\mu^\dagger(0)\psi(0) + {\rm h.c.},
\end{equation}
where  $\psi_\mu$ is an electron operator in the biased channel. In order to focus our analysis on the non-equilibrium LL effects, we consider these electrons to be effectively free, with the  local correlation function \eqref{biased}. 

Four different terms contribute to the correction \eqref{fpsi}. One can inject either a right- or a left-moving electron and collect either a right or a left mover. The asymptotic behavior of these contributions is summarized in Tab.\ \ref{tab:LL} for a  positive bias $\mu>0$.\footnote{Even though in the case of tunneling to a LL the time evolution in Eq.\ \eqref{corr_modified} cannot be trivially extended to infinity, the analysis similar to the one in Sec.\ \ref{sec:pt} can still be performed.}. For negative biases, $\mu<0$, one can use the electron-hole symmetry discussed in Sec.\ \ref{thissec} [see Eq.\ (\ref{e-h})]. 

\begin{table}[h!]
    
  \begin{center}
    \begin{tabular}{|c!{\vrule width 1pt}c|c|c|}
    \hline
     $\epsilon_0$ & $ -\mu+0$ & $ 0$ & $ \mu-0$\\
      \Xhline{1pt}
      R to R & $3\frac{K+K^{-1}}{2} + 1$ & $\frac{K+K^{-1}}{2} - 2$ & $\frac{K+K^{-1}}{2} - 1$\\
      \hline
      L to L & $3\frac{K+K^{-1}}{2} - 1$ & $\frac{K+K^{-1}}{2} - 2$ & $\frac{K+K^{-1}}{2} + 1$\\
      \hline
      L to R & $3\frac{K+K^{-1}}{2} $ & $\frac{K+K^{-1}}{2} - 2$ & $\frac{K+K^{-1}}{2} $\\
      \hline
    \end{tabular}
        \caption{The exponents $\kappa$ of the asymptotic behaviour $\delta n(\epsilon) \propto|\epsilon - \epsilon_0|^\kappa$ in the vicinity of different thresholds $\epsilon_0$ for tunneling to a LL are expressed in terms of the LL interaction parameter $K$. The results for different processes are listed, i.e., when left or right mover is injected and left or right mover is detected. The process R to L gives the same contribution as L to R.\label{tab:LL}} 
  \end{center}
\end{table}

\subsection{Fractional quantum Hall edge states}
\begin{figure}[h]
	\centering
		\includegraphics[scale=0.6]{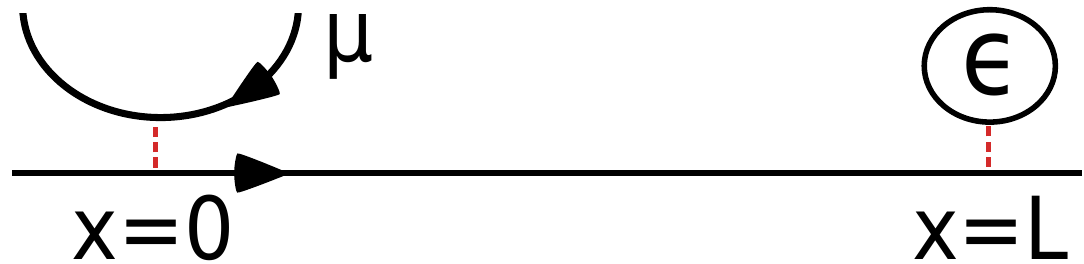} 
	\caption{A QH edge state at the  filling factor $\nu = (2n+1)^{-1}$, $n\in\mathbb{N}$, is schematically shown. Electrons  are injected at the point $x=0$ via the voltage biased QPC and detected at the point $x=L$ at energies $\epsilon$ with the help of a QD.}
\label{fig:FQH}
\end{figure}

Another interesting problem that cannot be solved by evaluating  the Fredholm determinant is the problem of the relaxation of a non-equilibrium stationary state at the edge of a fractional QH system. It is well known that at filling factors of the form $\nu = (2n+1)^{-1}$, $n\in\mathbb{N}$,  there exist a single channel of the free-propagating bosonic field $\phi$ at the edge\citep{fqh}. This field is related to the electron operator by the identity 
\begin{equation}
\label{bos_id}
\psi \propto e^{i\sqrt{2n+1}\phi}.
\end{equation}
We consider the system shown in Fig.\ \ref{fig:FQH}, where electrons tunnel between two fractional QH edges at the point $x=0$, and the relaxed stationary state is studied downstream at the point $x=L$. In this case, the tunneling Hamiltonian is given by
\begin{equation}
H_T = \tau \psi^\dagger_\mu \psi + {\rm h.c.},
\end{equation}
where $\psi_\mu$ is the fermion operator in the biased edge channel, and $\psi$ describes electrons  in the edge channel, where we study the correction $\delta n(\epsilon)$ to the stationary TDOS given by the Eq.\ \eqref{fpsi}. Note, that for the considered here fractional QH  system the correlator of electron fields in the biased channel takes the following form, $\langle\psi_\mu^\dagger(t)\psi_\mu(0)\rangle_{eq}\propto e^{i\mu t}/(it + 0)^{2n+1}$, which generalizes Eq.\ (\ref{biased}). 

The TDOS at the point $x=L$ is measured by a resonant tunneling through a QD level $\epsilon$. The results for $\mu>0$ are summarised in the Tab.\ \ref{tab:FQH}. The result for injecting holes, i.e., for $\mu<0$ can be obtained by using the symmetry $\delta n(-\epsilon)|_{\mu\to -\mu}=-\delta n(\epsilon)$. Note, that electron-like excitations $\mu>0$ do not affect the TDOS at negative energies ($\epsilon<0$). Technically, this follows from the fact that the expression under the integral in the Eq.\ \eqref{distr} becomes analytical, i.e., instead of integrals along the branch cuts one needs to compute residues of the poles. Another curious result is that at the threshold $\epsilon_0=0$ the exponent $\kappa$ vanishes. This implies that at these energies the correction may dominate over the background equilibrium TDOS, which vanishes at this point as $|\epsilon|^{2n}$ ($\epsilon<0$).
\begin{table}[h!]
  \begin{center}
    \begin{tabular}{|c!{\vrule width 1pt}c|c|c|}
    \hline
     $\epsilon_0$ & $ <0$ & $ +0$ & $ \mu-0$\\
      \Xhline{1pt}
      $\kappa$ & absent & $0$ & $2n$\\
      \hline
    \end{tabular}
        \caption{The exponents $\kappa$ of the asymptotic behaviour $\delta n(\epsilon)\propto|\epsilon - \epsilon_0|^\kappa$ in the vicinity of different thresholds $\epsilon_0$ for electron tunneling to a fractional QH edge at the filling factor $\nu = (2n+1)^{-1}$, $n\in\mathbb{N}$ are listed. Note, that the correction vanishes at negative energies $\epsilon<0$.  \label{tab:FQH}}
          \end{center}
\end{table}

Finally, we would like to mention, that various combinations of the electron and quasiparticle tunneling at the source and detector, as well as various other filling fractions,  can be experimentally relevant. They will be investigated elsewhere.


\section{Conclusion}
Exactly solvable strongly-interacting systems provide an important platform for studying the interplay between strong interaction and non-equilibrium physics. This is because it is possible to extend analytical results even beyond an equilibrium regime. With analytical predictions, one can test experimentally a current theoretical understanding both of interaction effects and of the non-equilibrium physics. In this paper we developed a new theoretical method that allows one to analyze strongly interacting systems out of equilibrium by studying an asymptotic universal behaviour of physical quantities in the vicinity of the thresholds in the spectrum of excitations. Our approach is based on the perturbation theory with respect to a small parameter, the number of non-equilibrium excitations, which is controlled by weak tunneling. We extended the results of previous works that use the Fredholm determinant technique to a class of systems without well defined electron excitations. Namely, in this paper along with conventional systems we studied the relaxation of non-equilibrium electrons in spinless LLs and at the edge of chiral fractional QH systems, where previously introduced methods cannot be applied.
 
\begin{acknowledgments}
This work has been supported by the Swiss National Science Foundation. We thank Anna Goremykina for useful comments and the critical reading of the manuscript. 
\end{acknowledgments}

\appendix
\section{Perturbative derivation of the electron correlation function from Fredholm determinant}
\label{sec:det}

In this Appendix we derive the Eq.\ \eqref{Kfast} from the Fredholm determinant representation of the electron correlation functions in a case when local tunneling process is effectively free-fermionic. We use the fact that the expression \eqref{feps2} can be reduced to a determinant of a single-particle operator by means of the non-equilibrium bosonization technique\citep{levk,ff2}. 
One of the key steps in this approach is to relate the bosonic fields at point $x=L$ to the transferred charge into the edge channel through the QPC at the point $x=0$. 

Since the eigenmodes propagate with constant speeds, one can write\citep{levk} 
\begin{align}
\phi_U(L,t) &= \alpha ^2 \phi_U(0,t-L/u_c)+\beta^2\phi_U(0,t-L/u_d)\nonumber\\
&+\alpha\beta \phi_D(0,t-L/u_d)-\alpha\beta \phi_D(t-L/u_c),
\end{align}
which allows one to present the electron correlator in terms of non-equilibrium correlators of bosonic field \eqref{chi_def}
\begin{align}
\label{detft}
\langle\psi_U^\dagger(L,t)&\psi_U(L,0)\rangle_{n-eq}  \\ \nonumber =
\chi_U&(t,\alpha^2)\chi_U(t,\beta^2)\chi_D(t,\alpha\beta)\chi_D(t,-\alpha\beta),
\end{align}
where we used the relations (\ref{bos_iden}) and (\ref{rotation}), and the simplification arising from the fact that dipole and charged wave packets are well separated in space in the $L\to \infty$ limit. Given the relation of the bosonic fields to the charge in the channel \eqref{bos_iden} one can express  $\chi_i(t,\lambda)$ in terms of the FCS of the charge $Q_i(t)$, $i=U,D$ transferred through the junction over time $t$
\begin{equation}
\label{FCS}
\chi_i(t,\lambda) = \langle e^{-2\pi i\lambda Q_i(t)}e^{2\pi i\lambda Q_i(0)}\rangle_{n-eq}
\end{equation}

In the case, where a local electron tunneling is effectively free fermionic, the evaluation of the correlator \eqref{FCS} amounts to solving the scattering problem at the source QPC and expressing the FCS generator in terms of the Fredholm determinant\citep{lev} (we use that $\log det = Tr\log$)
 \begin{equation}
\label{TR}
\log\chi_U(t,\lambda) = Tr\log\left(1-F+U_\lambda F\right),
\end{equation}
where $F$ is the diagonal in energy basis matrix with elements being the Fermi distribution functions in the incoming scattering channels. The matrix $U_\lambda$ is obtained from the scattering matrix (see Ref.\ [\onlinecite{lev}] for details) and is given by
\begin{equation}
U_\lambda=\begin{bmatrix}
1 & 0\\
0  & 1
\end{bmatrix}
+1_{0,t}\left(e^{i\lambda}-1\right)
\begin{bmatrix}
    T& rt^*\\
    r^*t & (1-T) 
\end{bmatrix},
\end{equation} 
where $r$ and $t$ are reflection and transmission amplitudes, respectively,  and $T = |t|^2$ is the tunneling probability. The role of this matrix is to ``count'' electrons, which end up in the channel of interest after scattering.

We  are interested in the limit of weak tunneling, therefore we can expand Eq.\ \eqref{TR} in small $T$. The zeroth order term $\log \chi_U^{(0)}(t,\lambda)$ represents the equilibrium charge fluctuations. The linear in $T$ contribution is given by
\begin{align}
&\log \chi_U^{(1)}(t,\lambda)\! \propto\! T\int_0^t\!\!\int_0^t dt_1 dt_2\left(\frac{t_1}{t-t_1}\right)^{\lambda}\left(\frac{t - t_2}{t_2}\right)^{\lambda} \nonumber \\
& \times e^{i\mu (t_1-t_2)}\left\{\frac{1}{(t_1-t_2-i0)^2}-\frac{e^{-2\pi i\lambda}}{(t_2-t_1-i0)^2}\right\}.
\end{align}
There are two contributions in Eq.\ \eqref{detft} that contain this term: one from the charged mode,  $\chi_U(t,\alpha^2)$ and another one from the dipole mode, $\chi_U(t,\beta^2)$. These are exactly two contributions that we have obtained in Sec.\ \ref{sec:pt}. The contribution of the charge mode to the correction \eqref{feps1} reads
\begin{equation}
\label{cor_from_det}
\delta n_c(\epsilon) \propto \int dt \frac{e^{-i\epsilon t}}{it+0}\log \chi_U^{(1)}(t,\alpha^2).
\end{equation}
This is nothing but Eq.\ \eqref{Kfast}, written in a different form. The contribution of the dipole mode $\delta n_d(\epsilon)$ is obtained by replacing $\alpha\to\beta$.

%

\bibliographystyle{apsrev4-1}
\bibliography{citation}

\end{document}